\documentclass[twocolumn]{aa}
\usepackage{graphicx}
\usepackage{txfonts}
\def \INT {{\it INTEGRAL}}
\def \ISGRI {{\it ISGRI}}
\def \EPIC {{\it EPIC}}
\def \XMM {{\it XMM}}
\def \src {IGR\,J16318$-$4848}
\def\approxgt{\mathrel{\hbox{\rlap{\lower.55ex \hbox {$\sim$}}
        \kern-.3em \raise.4ex \hbox{$>$}}}}
\def\approxlt{\mathrel{\hbox{\rlap{\lower.55ex \hbox {$\sim$}}
        \kern-.3em \raise.4ex \hbox{$<$}}}}

\begin{document}
   \title{\INT\ discovery of a bright highly obscured galactic X-ray binary source \src}

   \author{R. Walter\inst{1,2}, J. Rodriguez\inst{3,1}, L. Foschini\inst{4}, J. de Plaa\inst{5},
           S. Corbel\inst{3,6}, T. J.-L. Courvoisier\inst{1,2}, P. R. den Hartog\inst{5}
	   \\F. Lebrun\inst{3}, A. N. Parmar\inst{7}, J. A. Tomsick\inst{8}, P. Ubertini\inst{9} 
          }

\authorrunning{Walter R. et al.}
   \offprints{Roland.Walter@obs.unige.ch}

   \institute{\INT\ Science Data Centre, Chemin d'Ecogia 16, CH--1290 Versoix, Switzerland
         \and
             Observatoire de Gen\`eve, Chemin des Maillettes 51, CH-1290 Sauverny, Switzerland
         \and
             CEA Saclay, DSM/DAPNIA/SAp (CNRS FRE 2591), Bat 709, F--91191 Gif-sur-Yvette Cedex, France
         \and
	     IASF/C.N.R. Section of Bologna, Via Pietro Gobetti 101, I--40129 Bologna, Italy
         \and
	     SRON National Institute for Space Research, Sorbonnelaan 2, 3584 CA Utrecht, The Netherlands 
  	 \and
	     Universit\'e Paris VII (F\'ed\'eration APC), F--91191 Gif sur Yvette, France
         \and
	     Astrophysics Missions Division, Research and Scientific Support Department of ESA, ESTEC, PO Box 299, 2200 AG Noordwijk, The Netherlands
	 \and
	     CASS, Code 0424, University of California San Diego, La Jolla, CA 92093-0424, USA
         \and
             IASF/C.N.R. Section of Roma, Area di Ricerca di Tor Vergata, Via del Fosso del Cavaliere, I--00133 Roma, Italy
             }

   \date{Received July 14, 2003; accepted September 4, 2003}

   \abstract{
\INT\ regularly scans the Galactic plane to search for new objects and in particular for absorbed sources with the bulk of their emission above $10-20~{\rm keV}$. 
The first new \INT\ source was discovered on 2003 January 29, 0.5$\degr$ from the Galactic plane and was further observed in the X-rays with {\it XMM-Newton}. This source, \src, is intrinsically strongly absorbed by cold matter and displays exceptionally strong fluorescence emission lines. The likely infrared/optical counterpart indicates that \src\ is probably a High Mass X-Ray Binary neutron star or black hole 
enshrouded in a Compton thick environment. Strongly absorbed sources, not detected in previous surveys, could contribute significantly to the Galactic hard X-ray background between 10 and $200~{\rm keV}$.
}

   \maketitle
%

\section{Introduction}

X-ray binaries (where the compact object is a neutron star or black hole) can become strong hard X-rays emitters when accretion takes place. 
Among the $\sim$300 known X-ray binaries in our Galaxy and the Magellanic clouds, a few systems show strong intrinsic photo-electric absorption: GX 301--2 (Swank et al. 1976), Vela X--1 (Haberl \& White 1990), CI Cam (Boirin et al. 2002). Moderate absorption was also detected in a few X-ray bursters (Natalucci et al, 2000). We report here on the discovery of \src, a Compton thick X-ray binary in which the X-ray obscuring matter has a column density as large as the inverse of the Thomson cross section.

\section{High energy observations and data analysis}

\begin{figure*}
\centering
\includegraphics[bb=98 277 534 543,clip,width=16.0cm]{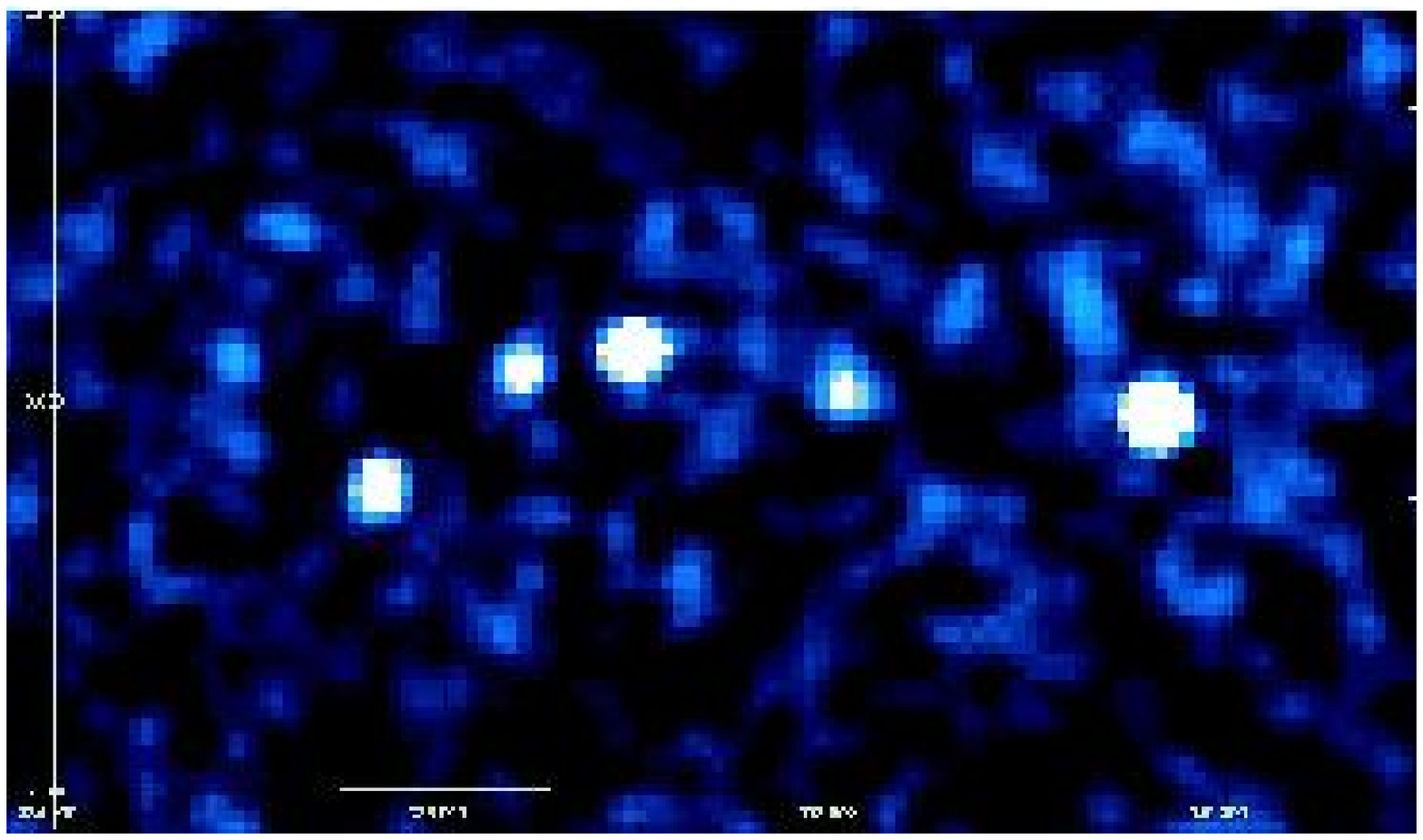}
\caption{\INT\ \ISGRI\ $15-40~{\rm keV}$ sky image of the Norma region in Galactic coordinates. This image accumulates 508~ksec of \INT\ core program data. The red circle indicates \src.}
\label{integral}
\end{figure*}

\src\ was discovered using the \INT\ imager {\it IBIS/ISGRI} (Ubertini et al. 2003; Lebrun et al. 2003) on 2003 January 29 (Courvoisier et al. 2003a) and was regularly observed for two months. Figure~\ref{integral} shows the $15-40~{\rm keV}$ \ISGRI\ sky image around the source with an accumulation time of $508~{\rm ksec}$. The position of the source ($RA = 16^{\rm h} 31.8^{\rm m}$ and $DEC = -48\degr 48'$) was determined with an accuracy of 2\arcmin. \src\ is detected up to $80~{\rm keV}$ with a mean $20-50~{\rm keV}$ flux of $6\times10^{-11}~{\rm erg~cm^{-2}~s^{-1}}$. Significant ($>5\sigma$) intensity variations occur on time scales as short as $1000~{\rm sec}$. 

\begin{figure*}
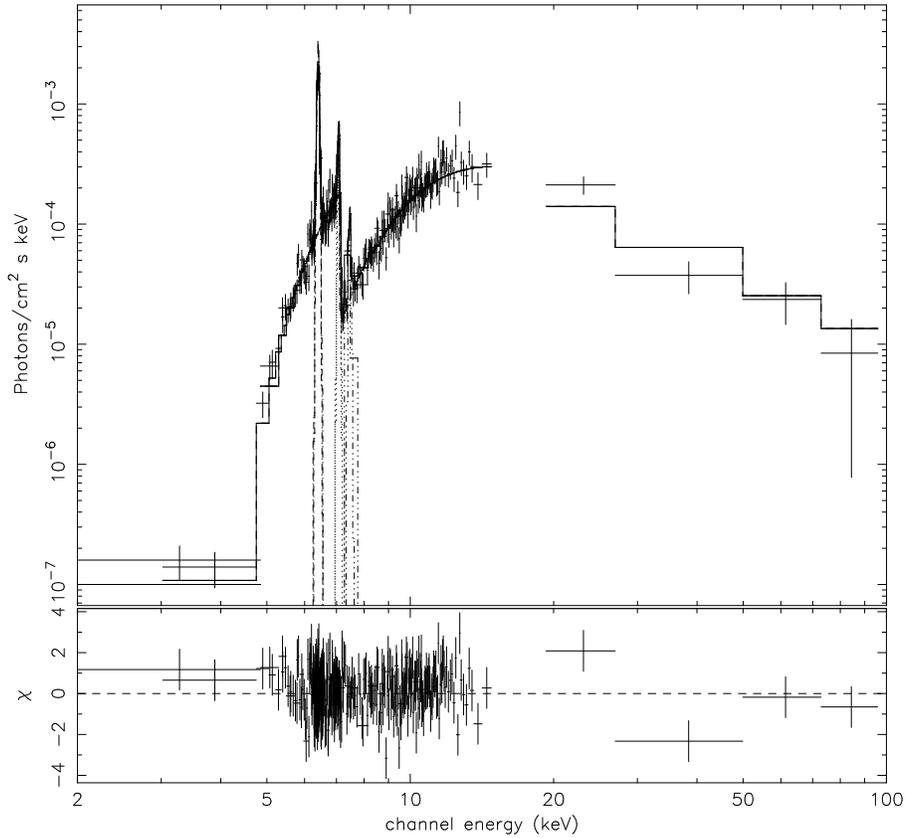

\centering
\includegraphics[bb=80 41 531 710,clip,angle=-90,width=11.9cm]{integral34_f2a.ps}
\includegraphics[bb=384 41 570 710,clip,angle=-90,width=11.9cm]{integral34_f2b.ps}
\caption{The {\it EPIC PN}, {\it EPIC MOS2} and \ISGRI\ photon spectra of \src\ along with the best fit model and residuals for $C_{\rm ISGRI}$ fixed to 0.67. The fit gave a reduced $\chi^2$ of 1.04 for 323 d.o.f. (table~1).}
\label{commonspec}
\end{figure*}

We selected 70 pointings of the \INT\ core program between revolution 36 and 50 in which \src\
was located at less than 5$^{\circ}$ from the center of the field of view. Since \src\ is often below the detection level, further selection was applied on the 
luminosity of the source. We selected pointings where the source was detected by the IBIS analysis software (Goldwurm et al. 2003)
and out of them only 8 pointings (from 2003 March 3 to 2003 March 14) for which we were able to produce spectra of sufficient quality. 
This resulted in a 17.5~ksec average spectrum that is displayed in Fig.~\ref{commonspec}. 
The current \ISGRI\ energy correction and response matrices are preliminary. Throughout this analysis we used an \ISGRI\ ancillary response that was modified to obtain a good fit to the spectrum of the Crab Nebula.
The \ISGRI\ spectrum of \src\ could be represented by a power law with a photon index, $\Gamma = 2.7^{+1.2}_{-0.8}$ and a flux ${\rm F_{20-100~{\rm keV}}} = 1.6\times10^{-10}~{\rm erg~cm^{-2}~sec^{-1}}$.

\src\ was observed for $28~{\rm ksec}$ by {\it XMM-Newton} on 2003 February 10. A single X-ray source was found within the \INT\ uncertainty circle in the {\it EPIC PN} and {\it MOS} cameras (Str\"uder et al. 2001; Turner et al. 2001) at a position of $RA = 16^{\rm h} 31^{\rm m} 48.6^{\rm s}$ and $DEC = -48\degr 49' 00"$ with a 4\arcsec\ uncertainty (Schartel et al. 2003). The X-ray spectrum was immediately recognized as exceptional featuring strong photo-electric absorption, the associated Fe absorption edge at $7.1~{\rm keV}$ and fluorescence line emission of mostly neutral Fe K$\alpha$ $(6.4~{\rm keV})$, Fe K$\beta$ $(7.1~{\rm keV})$ and Ni K$\alpha$ $(7.5~{\rm keV})$. The {\it XMM-Newton} spectrum of \src, presented by Matt \& Guainazzi (2003) is significantly flatter than the \ISGRI\ spectrum.

All spectral uncertainties are given at 90\% confidence for a single interesting parameter, unless indicated otherwise.
The abundances of Anders \& Grevesse (1989) and the photo-electric cross section of Verner et al (1996) are used throughout.

To analyze the \ISGRI\ and \EPIC\ data together we extracted \EPIC\ spectra using version 5.4.1 of the \XMM\ Science Analysis System (SAS) software. The data were first screened for enhanced variable background by filtering out the time intervals where the count rate above $10~{\rm keV}$ was higher than the threshold count rate (18 for {\it MOS2} and 60 for {\it PN} per $100~{\rm sec}$ bin). The total exposure
after screening resulted in $24~{\rm ksec}$ for {\it MOS2} and $21~{\rm ksec}$ for {\it PN}. Source events were subsequently extracted
from a 25$\arcsec$ radius circle centered on the source. A second circle with the same radius was fixed
on a comparable region on the detector to serve as background. Standard SAS tools were used to calculate the
instrumental response and the effective area for the extracted spectra. 

\begin{table}[h]
\caption{Best fit parameters (90\% confidence). All parameters are free in model 1. $C_{\rm ISGRI}$ is fixed to 0.67 in model 2}
\label{bestfit}
\begin{tabular}{lccl}
\hline\hline
Parameter          & Model 1             & Model 2       & Unit \\
\hline
$\chi^2$/d.o.f.    & 333 / 322           & 336 / 323       &                       \\
$C_{\rm ISGRI}$        &0.43$^{+0.14}_{-0.23}$& 0.67 (fixed)   &                       \\
$N_{\rm H}$   & 1.96 $\pm$ 0.07     & 2.07 $\pm$ 0.10 & $10^{24}~{\rm cm}^{-2}$\\      
$\Gamma$           & 1.6  $\pm$ 0.3      & 1.97 $\pm$ 0.17 &                       \\
Pwl $I_{1~{\rm keV}}$ & 0.08 $\pm$ 0.03     & 0.12 $\pm$ 0.03 & ${\rm ph/kev~cm^2~s}$\\
Fe$_{\rm abs}$         & 0.82 $\pm$ 0.05     & 0.79 $\pm$ 0.04 & ${\rm Z}_{\sun}$      \\
Fe K$\alpha$ E.    & 6.405$\pm$0.003     &6.405 $\pm$ 0.003& ${\rm keV}$           \\
Fe K$\alpha$ Flux  & 1.84 $\pm$ 0.09     & 1.75 $\pm$ 0.06 & $10^{-4}~{\rm ph/cm^2~s}$\\ 
Fe K$\beta$ E.     & 7.07 $\pm$ 0.01     & 7.07 $\pm$ 0.01 & ${\rm keV}$           \\
Fe K$\beta$ Flux   & 0.32 $\pm$ 0.08     & 0.30 $\pm$ 0.07 & $10^{-4}~{\rm ph/cm^2~s}$\\
Ni K$\alpha$ E.    & 7.46 $\pm$ 0.02     & 7.46 $\pm$ 0.02 & ${\rm keV}$           \\ 
Ni K$\alpha$ Flux  & 0.09 $\pm$ 0.03     & 0.09 $\pm$ 0.03 & $10^{-4}~{\rm ph/cm^2~s}$\\
Intr. line width   &$<$ 20               &$<$ 20           & ${\rm eV}$\\
\hline
\end{tabular}
\end{table}
The simultaneous fit of the {\it EPIC PN} and {\it MOS2} and \ISGRI\ spectra of \src\ using an absorbed power-law continuum, free Fe abundance and three Gaussian emission line model (model 1) resulted in a reduced $\chi^2$ of 1.04 for 322 degrees of freedom (d.o.f.). A normalization constant $C_{\rm ISGRI}$ was introduced in the model and was left as a free parameter in the fit in order to account for cross calibration uncertainties and the non simultaneity of the \ISGRI\ and \EPIC\ observations. The best fit parameters, listed in table~\ref{bestfit} are compatible with those of Matt \& Guainazzi (2003). The powerlaw flux in table 1 is corrected for the effects of photoelectric absorption but not for the effects of Compton scattering. The line fluxes are observed fluxes, not corrected for any absorption effect. 

Examination of the residuals of model 1 shows that the spectral model is systematically flatter than the \ISGRI\ spectrum. In addition as the \ISGRI\ spectrum represents the high state of the source and the \EPIC\ data correspond to a mix of different levels we would expect $C_{\rm ISGRI}$ to be $>0.67$, the value derived using our \ISGRI\ response matrix and the {\it EPIC/ISGRI} inter-calibration obtained on 3C273 (Courvoisier et al. 2003b), in contrast to the best fit value of 0.43. 

Matt \& Guainazzi (2003) noted that the spectral slope was weakly constrained by the \EPIC\ data alone as it correlates with the absorbing column density. $C_{\rm ISGRI}$ correlates with the power-law index as well. Constraining $C_{\rm ISGRI}$ to be $>0.67$ gives $\Gamma=1.9\pm0.2$ and a column density $N_{\rm H}=(2.1\pm0.1)\times10^{24}~{\rm cm}^{-2}$. The photon index is consistent with that derived from the fit to the \ISGRI\ spectrum alone. Table \ref{bestfit} gives the best fit parameter for a model with $C_{\rm ISGRI}$ fixed to 0.67 (model 2), the photon spectrum and the residuals are shown in Fig.~\ref{commonspec}. 

An exponential cutoff power-law continuum model also fits the data well providing a $\chi^2$ of 328 for 322 d.o.f. with a flat power-law and a cutoff energy of $15\pm5~{\rm keV}$. Simultaneous spectral observations performed below and above $20~{\rm keV}$ are necessary to constraint better the spectral model. 

The observed ratio of the intensities of Fe K$\beta$ and Fe K$\alpha$ ($0.17\pm0.09$) is consistent with the expected value of 0.14 (Kaastra \& Mewe 1993). The intensity ratio between Ni K$\alpha$ and Fe K$\alpha$ of $0.05\pm 0.02$ is close to the solar abundance ratio of 0.03-0.045 (Molendi et al, 2003). The strength of the Fe edge at $7.1~{\rm keV}$ also corresponds to that expected from a solar abundance. 

The centroids of the Fe K$\alpha$ and K$\beta$ lines corresponds to Fe that is ionized between 2 to 6 times ($1\sigma$ level) (House, 1969) and to a ionisation parameter $\Xi\approx0.05~{\rm erg~cm~s^{-1}}$ that is expected to be variable across the absorbing matter. The position of the Fe absorption edge also indicates the presence of Fe that is ionized less than 2 times. Note that the systematic uncertainty on the line and edge energies is $10~\rm{eV}$.

In contrast to Matt \& Guainazzi (2003), we did not find that the presence of a Compton shoulder to the Fe K$\alpha$ line was 
required by the data. This could be related to data selection and reduction. A firm detection of the Compton shoulder should be confirmed by future observations. The absence of detection of the Compton shoulder is however consistent with the conclusion of Matt \& Guainazzi that the average $N_{\rm H}$ on which fluorescence takes place could be smaller than that on the line of sight. We consider 
the flux of the Compton shoulder derived by Matt and Guainazzi (2003) as an upper limit.

We also did not find evidence for an excess of emission below $5~{\rm keV}$ which could be explained by the steeper continuum derived from the \INT\ data when compared with the use of XMM data alone. Our data do not therefore show evidence for reflection as could be inferred from the low energy excess. A pure transmission geometry, with in-homogeneously distributed absorbing mater is sufficient.

\begin{figure}[h]
\centering
\includegraphics[width=8.8cm]{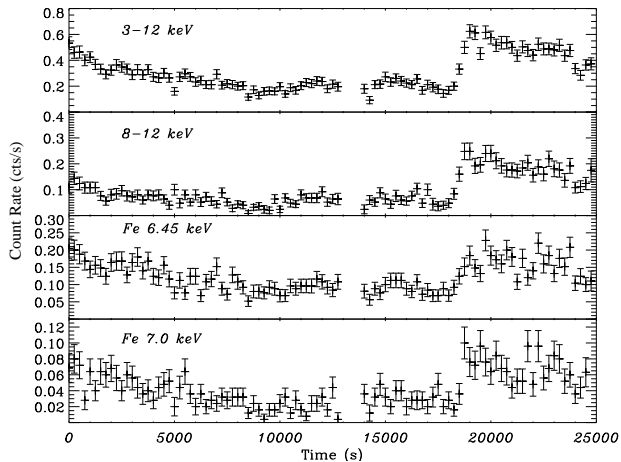}
\caption{{\it EPIC PN} light curves accumulated in different energy ranges for \src. The Fe K$\alpha$ light curve has been accumulated between 6.2 and $6.6~{\rm keV}$. The $8-12~{\rm keV}$ light curve is unaffected by line emission.}
         \label{XMMlc}
\end{figure}

Matt \& Guainazzi (2003) noted that the source flux variations were intrinsic to the source (and not related to absorption). We extracted X-ray light curves for the continuum and for the Fe K$\alpha$ line (Fig.~\ref{XMMlc}) using time bins of 200 sec. During the fast continuum rise (around time 19000 sec in Fig.~\ref{XMMlc}) the count rate varies significantly in 200 seconds and a cross correlation analysis between the $6.2-6.6~{\rm keV}$ light curve (strongly dominated by the Fe K$\alpha$ line) and the $8-12{\rm keV}$ continuum light curve indicates that any differences in mean arrival time of the continuum and Fe K$\alpha$ line emission are smaller than 200 sec. The ratio between the Fe K$\alpha$ and continuum shows however some significant variations which, as pointed out by Matt \& Guainazzi (2003), could indicate variations of the properties of the cold matter on time scales of $10^4~{\rm sec}$.

\section{Discussion}

\subsection{X-ray variability}

The variability time scale and the maximum delay observed between the Fe K$\alpha$ line and the continuum variations limit the size of the zone in which fluorescent emission takes place and its distance to the X-ray source to $10^{13}~{\rm cm}$. It is therefore very unlikely that \src\ is an extragalactic source such as a Seyfert II galaxy or an Ultra Luminous Infra Red Galaxy as the width of a fluorescence line (given by the Keplerian velocity) emitted at a distance of $10^{13}$ cm would be orders of magnitude larger than observed. 

Following the \INT\ discovery of \src, a re-analyis of archival data showed that \src\ had been weakly detected in 1994 September by ASCA (Murakami et al. 2003), with similar flux and $N_{\rm H}$ as observed in 2003. The ASCA spectrum suggested the presence of a strong Fe K$\alpha$ line (Revnivtsev et al. 2003). However, the {\it Beppo-SAX} Wide Field Camera (WFC), that observed the field almost continuously for 6 months each year between October 1996 and 2002, has never detected the source. This indicates that it was $\sim$10 times fainter on average during those periods (In't Zand 2003). This suggests that \src\ was active in 1994 and 2003 and that it has been quiet over periods of many months. It is remarkable that the flux and the absorption observed in the active states in 1994 and 2003 are very similar.

\subsection{Counterpart}

\begin{figure}
\centering
\includegraphics[bb=35 260 510 577,clip,width=8.5cm]{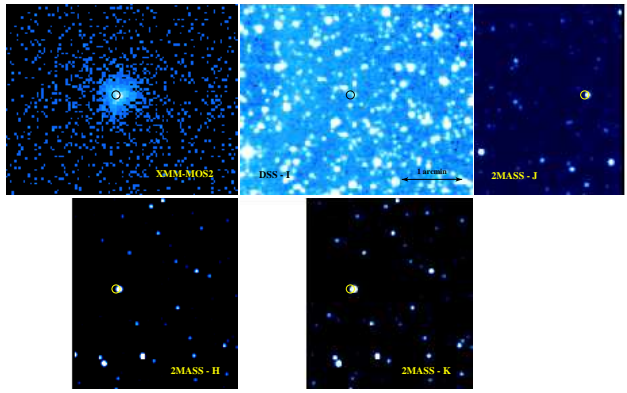}
\caption{{\it EPIC MOS}, {\it 2MASS} (J, H, K) and {\it DSS} images of the counterpart of \src. All images have the same scale. The \XMM\ error circle (4\arcsec\ radius) is shown on each image.}
\label{counterpart}
\end{figure}

A possible counterpart to \src\ in the Digitized Sky Survey (DSS--II/USNO-B1.0), Two Microns All Sky Survey (2MASS), and Midcourse Space Experiment (MSX) data was reported by Foschini et al. (2003). The respective images are shown in Fig.~\ref{counterpart}. In the 2MASS, the counterpart has $J=10.2$, $H=8.6$, $K=7.6$ and an uncertainty of $\pm0.3~{\rm mag}$. It is also clearly detected in the I band of the second DSS and also in the R band (USNO-B1). The R magnitude is reported to vary over an interval of 50 years between $17.3\pm0.3$ and $18.4\pm0.3$. The flux density in the MSX A band is $0.46~{\rm Jy}$.   

Radio observations were performed with the Australia Telescope Compact Array (ATCA). Observations have been conducted at 4.8 and $8.6~{\rm Ghz}$ (with a total bandwidth of 128 MHz) on 2003, February 9, starting at 18:00 and finishing at 04:00 the following day, with a total of 1.33 hours (spread over the full ATCA run) on \src. No source was detected with a $1~\sigma$ upper limit of 0.1 mJy both at 4.8 and $8.6~{\rm GHz}$.

The near infrared spectral energy distribution corrected for the effect of various possible values of the galactic absorption (Lutz et al. 1996) was analyzed. The visual extinction factor derived from the galactic $N_{\rm H}$ along the line of sight is $A_{\rm V} = 11$. The absorption could however be smaller if the source is nearby or larger as the radio measurements have limited spatial resolution. The maximum reddening compatible with an infrared spectrum not flatter than a black body spectral component $(A_{\rm V} = 20)$ is two orders of magnitude smaller than the absorption observed in the X-rays. We conclude that the source of the IR/optical emission is located at $>10^{13}~{\rm cm}$ from the X-ray source. 

If the reddenning is low $({\rm A_V}\le11)$ the infrared companion of \src\ is a low mass red giant star  with a luminosity of $100~d_{\rm kpc}^2~{\rm L}_{\sun}$. If the reddening is strong $({\rm A_V}\approx 20)$ the companion is a massive supergiant star of luminosity $10^5~d_{5 {\rm kpc}}^2~{\rm L}_{\sun}$. In the rest of this discussion we will assume that the infrared/optical counterpart and \src\ are a binary system. This however remains to be verified.

If the system is a Low Mass X-Ray Binary ($LMXRB$) located at 1~kpc, the unabsorbed $20-50~{\rm keV}$ luminosity of $3\times 10^{34}~{\rm erg~s}^{-1}$ corresponds to the quiescent state of those systems. In that state, neutron star systems emit thermal emission $({\rm kT}\approx 100~{\rm eV})$ that dominates the hard tail below $3~{\rm keV}$ (Rutledge et al. 2002) which is not detected in \src. Black hole $LMXRB$ systems in quiescence have lower luminosities (Kong et al. 2002) and do not show strong absorption.

Alternatively, if \src\ is a High Mass X-ray Binary ($HMXRB$) located at 5~kpc, the unabsorbed $20-50~{\rm keV}$ luminosity of $7\times10^{35}~{\rm erg~s}^{-1}$ indicates that moderate (wind) accretion is taking place. Strong Fe K$\alpha$ lines have been observed in other $HMXRB$ systems. In Vela X-1, large line equivalent widths were observed during eclipses when only scattered X-rays are observed, the absorbing $N_{\rm H}$ remained however smaller than observed in \src\ (Pan et al. 1994). In GX 301--2, the $N_{\rm H}$ and the equivalent width of the Fe K$\alpha$ line are variable and correlated. White \& Swank (1984) explained this behavior by variations of the stellar wind velocity and density along the orbit of the compact source. The Fe K$\alpha$ equivalent width and the $N_{\rm H}$ observed from \src\ match the extreme of the correlation found in GX 301--2. This suggests that the stellar wind accreting onto the compact source of \src\ could form a dense spherical shell in which fluorescence and absorption takes place. This possibility was also proposed by Revnivtsev et al. (2003). Note that the exponential cutoff powerlaw model that was used to represent the data is rather typical for accreting X-ray pulsars.

The ionisation parameter derived from our observations can be used to estimate the distance between the fluorescing material and the X-ray source $D=(a/D)~L/\Xi~N_{\rm H}=10^{13}~(a/D)~{\rm cm}$ where $a$ is the shell thickness. This distance is also compatible with the line variability. This distance compares better with the companion star radius than with the accretion radius (unless the stellar wind velocity is small).

We searched unsuccesfully for the presence of pulsations that would be a clear signature of a neutron star in the system. The counting rate of the source during the XMM observation does not allow any definitive answer as the upper limit on the relative amplitude for a rather broad 0.15 Hz quasi periodic oscillation (QPO) (e.g. $\nu_{\rm centroid}/{\rm FWHM}~=~3)$ is $\sim 30 \%$ at the 3 $\sigma$ confidence level, which is not constraining. An $8000~{\rm sec}$ observation with the Proportional Counter Array (PCA) onboard {\it RXTE} did not find evidence for QPO either (Swank \& Markwardt, 2003).

\subsection{Contribution to the X-ray background}

Absorbed sources could contribute significantly to the Galactic high energy background. Valinia et al. (2000a) modeled the Galactic background emission observed by the PCA on {\it RXTE} and OSSE on {\it CGRO} using a specific spectral component dominating between 10 and $200~{\rm keV}$. The flux of that component $(2.5\times10^{-11}~{\rm erg~cm^{-2}~s^{-1}~deg^{-2}}$ at $40~{\rm keV})$ is variable indicating that an important fraction of that emission comes from point sources. The spectral shape of \src\ roughly correspond to the empirical spectral model used to represent the Galactic background. The intensity of the background emission could be explained by 0.4 sources per square degree at the level of \src\ in a band of a width of few degrees around the Galactic plane.

The hard X-ray background also displays a strong Fe K$\alpha$ emission line which is attributed to the thermal diffuse emission dominating below $10~{\rm keV}$ (Valinia et al. 2000b). A population of intrisically absorbed sources similar to \src\ would only contribute at a level of 10\% to the background line intensity.

In spite of the {\it BeppoSAX} WFC long term monitoring of the Galactic center only few strongly absorbed sources have been discovered so far (Ubertini et al. 1999). \INT\ is able to make sensitive search for highly absorbed sources (Lebrun et al. 1999) such as \src, IGR\,J16320$-$4751 (Rodriguez et al. 2003) and IGR\,16358$-$4726 (Revnivtsnev et al. 2003). Figure~\ref{integral} shows the detection by \INT\ of one point source per degree of Galactic longitude in the Norma region. Those sources could indeed explain a significant fraction of the Galactic diffuse emission.

\begin{acknowledgements}
This work is based on observations obtained with \INT\ and {\it XMM-Newton}, two ESA science missions with instruments, science data centre and contributions funded by the ESA member states with the participation of the Czeck Republic, Poland, Russia and the USA. 

We thank S. Chaty, M. Del Santo, R. Fender, W. Hermsen, J.S. Kaastra, P. Laurent, M. Mendez, T. Tzioumis, J.J.M. In't Zand and J. Zurita. JR acknowledges financial support from the French Space Agency (CNES). LF acknowledges the hospitality of the ISDC during part of this work and the financial support from the Italian Space Agency (ASI).

This publication makes use of data products from the Two Micron All Sky Survey, which is a joint project of the University of Massachusetts and the Infrared Processing and Analysis Center/California Institute of Technology, funded by the NASA and the National Science Foundation. 

This research has made use of The Digitized Sky Surveys that were produced at the Space Telescope Science Institute under U.S. Government grant NAG W-2166.

\end{acknowledgements}

\end{document}